# Invariant Eigen-Structure in Complex-Valued Quantum Mechanics

**Ciann-Dong Yang, Shiang-Yi Han**

*Department of Aeronautics and Astronautics, National Cheng Kung University, Taiwan*

*Email: cdyang@mail.ncku.edu.tw*

**Abstract**

The complex-valued quantum mechanics considers quantum motion on the complex plane instead of on the real axis, and studies the variations of a particle's complex position, momentum and energy along a complex trajectory. On the basis of quantum Hamilton-Jacobi formalism in the complex space, we point out that having complex-valued motion is a universal property of quantum systems, because every quantum system is actually accompanied with an intrinsic complex Hamiltonian originating from the Schrödinger equation. It is revealed that the conventional real-valued quantum mechanics is a special case of the complex-valued quantum mechanics in that the eigen-structures of real and complex quantum systems, such as their eigenvalues, eigenfunctions and eigen-trajectories, are invariant under linear complex mapping. In other words, there is indeed no distinction between Hermitian systems, PT-symmetric systems, and non PT-symmetric systems when viewed from a complex domain. Their eigen-structures can be made coincident through linear transformation of complex coordinates.

**Keywords:** Complex Hamiltonian; Complex Bohmian mechanics; Intrinsic Hamiltonian

## 1. Introduction

Hermiticity as a critical property providing the reality of spectrums had been replaced by a weaker property called PT-symmetry [1]. Recently, complex spectrums in PT-symmetric quantum mechanical systems have brought a resurgence of interest in discussing the complex extension [2]. It brings effects on the discussion of complex properties of quantum mechanical systems, and gains attention to the exploration of the reality reflected by complex features of the microscopic world [3,4]. Some classical approaches to those complex issues have been proposed [5,6], in which they embed classical trajectories in complex space. The necessity of extending quantum mechanics to a complex domain comes from the observation that the Schrödinger equation can be transformed into the quantum Hamilton-Jacobi (H-J) equation, wherein canonical variables are complex-valued [7]. Based on the quantum H-J equation, quantum Hamilton mechanics was established in [8,9], where a fully classical interpretation of quantum mechanics was proposed in a complex domain. It was pointed out that the wave-particle duality, the uncertainty principle, and the Feynman's multi- path phenomenon all originate from the projection effect from the complex space to the real space [10-12]. Strong evidences from the El Naschie's $E$-infinity theory [13,14] and the many other accompanied works [15-17] further indicate that quantum phenomena are produced by the non-classical topology and geometry of quantum spacetime when projected into our 3+1 Euclidean space [18-20].

By using the complex-space formulation of fractal spacetime [21-22], many quantum phenomena, such as tunneling [23], wave-particle duality [10], spin [24,25], state transition [26], path integral [12], quantum chaos [27], uncertainty principle [11] and molecular dynamics [28,29], have been modeled exactly in the framework of complex-extended Hamilton mechanics, wherein quantum motions are described by complex-valued nonlinear differential equations. Apparently, quantum motions within complex space provide us with a better understanding of the quantum world in terms of a deterministic interpretation than a probabilistic one which is the only vision given by quantum mechanics.

It is revealed in this paper that having complex-valued Hamiltonians is a universal property of quantum systems, because every quantum system is actually accompanied with an intrinsic complex Hamiltonian originating from the Schrödinger equation. Quantum Hamilton mechanics, as formulated in complex space, is an ideal tool for studying different kinds of complex Hamiltonians, such as PT-symmetric Hamiltonians, non PT-symmetric Hamiltonians and other form of non-Hermitian Hamiltonians. We will see that quantum systems subjected to different complex potentials, having totally different real or imaginary eigenvalues, may possess similar intrinsic complex Hamiltonians and generate similar trajectories. Contrary to the common belief, we point out here that there is indeed no distinction between Hermitian system, PT-symmetric system, and non PT-symmetric system when they are viewed from the complex domain. Their quantum trajectories and eigenfunctions can be made identical under linear coordinate transformation in the complex plane.

In the following sections, we first introduce quantum Hamilton mechanics in Section 2 and show that every quantum system accompanies an intrinsic complex Hamiltonian. This intrinsic complex Hamiltonian represents energy conservation, and is unique to the quantum system. In Section 3, an objective physical process dealing with a harmonic oscillator system will be introduced on the basis of quantum Hamilton mechanics.

Two non-Hermitian Hamiltonians, which are deformations of a harmonic oscillator, will be considered in Section 4 to show that Hermitian and non-Hermitian quantum systems can be connected via linear coordinate mapping in the complex plane. Then in Section 5, we propose a general Hamiltonian, which comprises all complex Hamiltonians, either PT- symmetric or not, that can be made identical to the Hermitian Hamiltonian of a harmonic oscillator by linear mapping of complex coordinates.

As the last topic discussed in this paper, the physical meaning of complex energy in reality will be introduced. We reveal three possible sources of the complex energy and numerical results are illustrated to support our notion about the complex energy. The final section are discussions and conclusions.

## 2. Intrinsic Complex Hamiltonian

Consider a quantum mechanical system described by a wavefunction $\Psi(t,\mathbf{q})$. To derive the intrinsic complex Hamiltonian associated with $\Psi(t,\mathbf{q})$, we first express $\Psi(t,\mathbf{q})$ in terms of a complex action function $S(t,\mathbf{q})$:

$$\Psi(t,\mathbf{q}) = e^{iS(t,\mathbf{q})/\hbar}, \qquad (2.1)$$

through which we can transform the Schrödinger equation

$$i\hbar \frac{\partial \Psi}{\partial t} = -\frac{\hbar^2}{2m}\nabla^2 \Psi + V\Psi, \qquad (2.2)$$

into the quantum H-J equation

$$\frac{\partial S}{\partial t} + \left[\frac{\mathbf{p}^2}{2m} + V(t,\mathbf{q}) + \frac{\hbar}{2m\mathrm{i}}\nabla \cdot \mathbf{p}\right] = 0, \qquad (2.3)$$

where $\mathbf{p} = \nabla S$ is the complex canonical momentum associated with $\mathbf{q}$. The quantum Hamiltonian $H$ now reads from Eq. (2.3) as

$$H(\Psi) = \frac{1}{2m}\mathbf{p}^2 + V(t,\mathbf{q}) + Q(\Psi). \qquad (2.4)$$

In addition to the first two classical components in Eq. (2.4), the complex Hamiltonian comprises a new term known as the complex quantum potential:

$$Q(\Psi) = \frac{\hbar}{2m\mathrm{i}}\nabla \cdot \mathbf{p} = -\frac{\hbar^2}{2m}\nabla^2 \ln \Psi(t,\mathbf{q}). \qquad (2.5)$$

The complex quantum potential $Q(\Psi)$ is uniquely determined by the wave function $\Psi$ and is intrinsic in quantum systems. The one-to-one correspondence between $H(\Psi)$ and $\Psi$ allows us to interpret the probability information contained in $\Psi$ in terms of the nonlinear Hamilton dynamics contained in $H(\Psi)$. The Hamilton equations of motion can be derived from $H(\Psi)$ as

$$\dot{\mathbf{q}} = \frac{\partial H(\Psi)}{\partial \mathbf{p}} = \frac{\mathbf{p}}{m}, \qquad (2.6)$$

$$\dot{\mathbf{p}} = -\frac{\partial H(\Psi)}{\partial \mathbf{q}} = -\frac{\partial(V+Q)}{\partial \mathbf{q}}. \qquad (2.7)$$

The complex action function $S(t,\mathbf{q})$ brings us a complex canonical momentum,

$$\mathbf{p} = \nabla S(t,\mathbf{q}) = -\mathrm{i}\hbar \nabla \ln \Psi(t,\mathbf{q}), \qquad (2.8)$$

in which the generalized coordinate $\mathbf{q}$ as a function of time $t$ is complex-valued and can be found by integrating Eq. (2.6) with $\mathbf{p}$ given by Eq. (2.8),

$$m\dot{\mathbf{q}} = \nabla S(t,\mathbf{q}) = -\mathrm{i}\hbar \nabla \ln \Psi(t,\mathbf{q}), \qquad (2.9)$$

which might be conceived as a complex extension of the de Broglie-Bohm guidance condition [30,31]. In fact, the complex nature of $(\mathbf{q}(t),\mathbf{p}(t))$ inherits from their associated quantum operators. For instance, by rewriting Eq. (2.9) as $-i\hbar\nabla\Psi = \mathbf{p}\Psi$ and comparing it with the definition of the momentum operator, $\hat{\mathbf{p}}\Psi = \mathbf{p}\Psi$, it naturally leads to the quantization axiom $\hat{\mathbf{p}} = -i\hbar\nabla$ [8,32].

The combination of Eq. (2.6) and Eq. (2.7) leads to the Newton's equation in the complex domain:

$$m\frac{d^2\mathbf{q}}{dt^2} = -\frac{\partial(V+Q)}{\partial\mathbf{q}} = -\frac{\partial V_{\text{Total}}(t,\mathbf{q})}{\partial\mathbf{q}}, \qquad (2.10)$$

which explicitly characterizes the interaction between the externally applied force $-\partial V/\partial\mathbf{q}$ and the intrinsic quantum force $-\partial Q/\partial\mathbf{q}$ in the quantum state $\Psi$.

When the applied potential $V$ is time-independent, $V = V(\mathbf{q})$, the wavefunction can be expressed by $\Psi(\mathbf{q},t) = \psi(\mathbf{q})e^{-i(E/\hbar)t}$ and its accompanying action function become

$$S(\mathbf{q},t) = -i\hbar\ln\psi(\mathbf{q}) - Et \qquad (2.11)$$

The quantum H-J equation (2.3) is then reduced to an energy conservation law:

$$H(\mathbf{q},\mathbf{p},t) = \frac{\mathbf{p}^2}{2m} + V(\mathbf{q}) + Q(\Psi(\mathbf{q})) = -\partial S/\partial t = E = \text{const.} \qquad (2.12)$$

With momentum $\mathbf{p}$ given by Eq. (2.8), Eq. (2.12) is further reduced to the time-independent Schrödinger equation

$$\frac{\hbar^2}{2m}\nabla^2\Psi(\mathbf{q}) + (E - V(\mathbf{q}))\Psi(\mathbf{q}) = 0. \qquad (2.13)$$

It reveals that the essential of the Schrödinger equation indeed represents the energy conservation in a microscopic world.

## 3. Intrinsic Complex Hamiltonian of a Harmonic Oscillator

Consider the classical Hamiltonian for a one-dimensional harmonic oscillator in dimensionless form with $\hbar = m = k = 1$:

$$H_1^{(c)} = (p_1^2 + x_1^2)/2. \qquad (3.1)$$

The corresponding quantum mechanical system is obtained by the quantization axiom:

$$H_1\psi_1 = \left(p_1^2 + x_1^2\right)\psi_1 = (-d^2/dx_1^2 + x_1^2)\psi_1 = E_1\psi_1, \qquad (3.2)$$

from which eigenfunctions $\psi_{1,n}$ and eigen-values $E_{1,n}$ are found to be

$$\psi_{1,n}(x_1) = C_n H_n(x_1) e^{-x_1^2/2}, \qquad (3.3)$$

$$E_{1,n} = n + 1/2, \; n = 0, 1, \cdots, \qquad (3.4)$$

where $H_n(x_1)$ is the $n^{\text{th}}$-order Hermite polynomial. Although $H_1$ is obtained from $H_1^{(c)}$, this does not mean that we may equate $H_1^{(c)}$ to $E_{1,n}$ and conclude that the classical Hamiltonian $H_1^{(c)}$ is stationary at energy levels specified by $E_{H,n}$.

However, if we replace $H_1^{(c)}$ by the complex Hamiltonian $H_1$ which is intrinsic in the quantum system defined by Eq. (2.4), then the previous statement becomes true. In other words, with $\psi_n$ given by Eq. (3.3), we can verify from Eq. (2.12) the following energy conservation law:

$$H_1(x_1) = \frac{1}{2}p_1^2 + \frac{1}{2}x_1^2 - \frac{1}{2}\frac{d^2 \ln \psi_{1,n}}{dx_1^2} = E_{1,n} = n + 1/2, \qquad (3.5)$$

where the dimensionless momentum $p_1$ is given by Eq. (2.8) as

$$\dot{x}_1 = \frac{\partial H_{1,n}}{\partial p_1} = p_1 = \frac{1}{i}\frac{d \ln \psi_{1,n}(x_1)}{dx_1}. \qquad (3.6)$$

It can be seen that the coordinate $x_1$ can only be defined in the complex plane for Eq. (3.6) to have a solution. Each eigenstate $\psi_{1,n}$ has its own eigen-trajectory $x_1(t)$ determined by Eq. (3.6). For instance, the equation of motion for the ground state, $n = 0$, can be obtained by substituting $\psi_{1,0}(x_1) = e^{-x_1^2/2}$ into Eq. (3.6)

$$\dot{x}_1 = ix_1, \quad x_1(0) \in \mathbb{C}. \qquad (3.7)$$

The resulting ground-state eigen-trajectory, $x_1(t)$, is illustrated in Fig.1. Eq. (3.7) has the solution $x_H(t) = Ae^{it}$, which, when expressed in the physical units, becomes

$$x_1(t) = A\cos\omega t + \mathrm{i}A\sin\omega t, \tag{3.8}$$

where $\omega = \sqrt{k/m}$ is the oscillation frequency. It is noteworthy that the ground-state quantum motion $x_1(t)$ is identical to the classical motion governed by $m\ddot{x} + kx = 0$.

Similarly, the equation of motion for the first excited state is given by,

$$\dot{x}_1 = \mathrm{i}(x_1^2 - 1)/x_1. \tag{3.9}$$

Several typical eigen-trajectories for the $n=1$ state are shown in Fig.2. There are two equilibrium points $x_{Heq} = \pm 1$ in this nonlinear dynamics and three sets of eigen-trajectories $\Omega_1$, $\Omega_2$, and $\Omega_3$ can be identified.

In addition, closed trajectories in the complex plane permit us to acquire the periods of oscillation by using the Cauchy theory. It provides us quantized periods in all states. In the $n=1$ state, periods are evaluated from the contour integration of Eq. (3.9) along any closed complex contour $c$ traced by the particle

$$T_1 = \frac{1}{\mathrm{i}}\int_c \frac{x_1}{x_1^2 - 1}\,dx_1 = \begin{cases} \pi, & c \in \Omega_1, \Omega_2, \\ 2\pi, & c \in \Omega_3. \end{cases} \tag{3.10}$$

All the trajectories encircling only one of the equilibrium points thus have the same period $T = \pi\sqrt{m/k}$, and those encircling both of the equilibrium points have period $T = 2\pi\sqrt{m/k}$. In fact, all quantization rules can be validated via the evaluation of a contour integral along any eigen-trajectory [33]. One of the most significant quantization rules was proposed by Sommerfeld and Wilson,

$$\oint p\,dx = \hbar\oint p\,dx = nh, \quad n = 0, 1, \cdots. \tag{3.11}$$

According to Eq. (3.9), we can evaluate the contour integration of $p_1$ along the eigen-trajectory $x_1(t)$ to obtain

$$\oint_c p_1 dx_1 = i\oint_c \frac{x_1^2 - 1}{x_1} dx_1 = \begin{cases} 2\pi, & \forall c \text{ encloses origin} \\ 0, & \text{otherwise} \end{cases} \quad (3.12)$$

These results are compatible with the quantization rule (3.11) and imply that along all possible closed eigen-trajectories in the $n = 1$ eigenstate, the action variable $J = \oint p_1 dx_1$ is quantized to two discrete values: $0$ or $h$. This quantization phenomenon can also be verified graphically by the sets of eigen-trajectories, $\Omega_1$, $\Omega_2$, and $\Omega_3$ indicated in Fig.2. Since $\Omega_1$ and $\Omega_2$ do not enclose the origin and thus give zero action variable $J$; while $\Omega_3$ encloses the origin and the residue theorem gives $J = h$ in physical units. Three sets of eigen-trajectories can be found: $\Omega_1$ and $\Omega_2$ contain all the trajectories encircling $x_{eq} = -1$ and $x_{eq} = 1$, respectively. $\Omega_3$ contains all the trajectories encircling both $x_{eq} = \pm 1$.

## 4. Non-Hermitian Hamiltonian of a Harmonic Oscillator

Particle motions in non-Hermitian systems will be addressed in this section. Non-Hermitian systems may be either PT-symmetric or non-PT-symmetric. A PT-symmetric system has complex eigenfunctions but its eigenvalues are still real. A non-PT-symmetric system has complex eigenvalues and complex eigen-functions. Consider a typical PT-symmetric system $H_2^{(c)}$ and a non-PT-symmetric system $H_3^{(c)}$ represented by

$$H_2^{(c)} = (p_2^2 + x_2^2 + ix_2)/2, \quad (4.1)$$

$$H_3^{(c)} = (p_3^2 + x_3^2 + ix_3 - x_3)/2. \quad (4.2)$$

From the viewpoint of quantum mechanics, the complexity of the Hamiltonian in Eq. (4.1) is only due to the multiplication by the imaginary number $i$, since $p_2$ and $x_2$ are treated as real variables therein; while in quantum Hamilton mechanics introduced here, both $p_2$ and $x_2$ are complex variables, and the Hamiltonian is complex-valued, regardless of the appearance of the $i$ factor.

Defining canonical variables $(x, p)$ in a complex plane has a remarkable significance that Hermitian systems such as $H_1^{(c)}$, PT-symmetry systems such as $H_2^{(c)}$, and non-PT-symmetry systems such as $H_3^{(c)}$, can be unified into a more general class of Hamiltonians whose elements are all linear-mapping invariant over the complex plane. The eigenfunctions for $H_2$ and $H_3$ can be found as

$$\psi_{2,n}(x_2) = C_n H_n\left(x_2 + i/2\right) e^{-(x_2+i/2)^2/2}, \tag{4.3a}$$

$$\psi_{3,n}(x_3) = C_n H_n\left(x_3 + \frac{i}{2} - \frac{1}{2}\right) e^{-\left(x_3 + \frac{i}{2} - \frac{1}{2}\right)^2/2} \tag{4.3b}$$

In terms of the Hermitian eigenfunction $\psi_{1,n}$ in Eq. (3.3), we can express $\psi_{2,n}$ and $\psi_{3,n}$ as

$$\psi_{2,n}(x_2) = \psi_{1,n}(x_1)\big|_{x_1 = x_2 + i/2}, \tag{4.4a}$$

$$\psi_{3,n}(x_3) = \psi_{1,n}(x_1)\big|_{x_1 = x_3 + i/2 - 1/2}. \tag{4.4b}$$

which means that the three eigenfunctions $\psi_{1,n}$, $\psi_{2,n}$ and $\psi_{3,n}$ can be made identical by linear coordinate translation over the complex plane. This relation also reflects in their respective intrinsic complex Hamiltonian:

$$H_2(x_2) = \frac{1}{2}\left[p_2^2 + \left(x_2 + \frac{i}{2}\right)^2 - \frac{d^2}{dx_2^2}\ln\psi_{2,n}(x_2)\right] + \frac{1}{8} \tag{4.5a}$$

$$H_3(x_3) = \frac{1}{2}\left[p_3^2 + \left(x_3 + \frac{i}{2} - \frac{1}{2}\right)^2 - \frac{d^2}{dx_3^2}\ln\psi_{3,n}(x_3)\right] + \frac{i}{4} \tag{4.5b}$$

Inserting relations (4.4) into Eqs. (4.5) and comparing them with the standard intrinsic complex Hamiltonian $H_1(x_1)$ in Eq. (3.5), we then have

$$H_2(x_2) = H_1(x_1)\big|_{x_1 = x_2 + i/2} + 1/8, \tag{4.6a}$$

$$H_3(x_3) = H_1(x_1)\big|_{x_1=x_3+i/2-1/2} + i/4, \tag{4.6b}$$

which shows that there is a constant real energy shift, $1/8$, between $H_2(x_2)$ and $H_1(x_1)$, and a constant imaginary energy shift, $i/4$, between $H_3(x_3)$ and $H_1(x_1)$. According to the energy conservation law (3.5), the energy levels for $H_2$ and $H_3$ in Eqs. (4.6) can be expressed in terms of $E_{1,n}$ as

$$E_{2,n} = E_{1,n} + 1/8 = n + 5/8, \tag{4.7a}$$

$$E_{3,n} = E_{1,n} + i/4 = n + 1/2 + i/4. \tag{4.7b}$$

The same results can also be obtained by calculating the eigenvalues of $H_2$ and $H_3$ directly. The results of Eqs. (4.7) confirm the prediction that the PT-symmetric system $H_2$ has real eigenvalues, while the non-PT-symmetric system $H_3$ has complex eigenvalues.

In the usual interpretation, $x_1$, $x_2$, and $x_3$ are regarded as real variables, and consequently, the eigenfunction $\psi_{1,n}(x_1)$ in Eq. (3.3) is regarded as real because $x_1$ is real, but the eigenfunctions $\psi_{2,n}(x_2)$ and $\psi_{3,n}(x_3)$ in Eqs. (4.3) are complex due to the appearance of the imaginary number $i$. This is the very reason why the distinctions between Hermitian systems, PT-symmetry systems, and non-PT-symmetry systems can be brought out. The system $H_1$ is said to be Hermitian, because both its eigenfunction $\psi_{1,n}(x_1)$ and eigenvalue $E_{1,n}$ are real. The eigenfunction $\psi_{2,n}(x_2)$ of $H_2$ is complex, but its eigenvalue $E_{2,n}$ is real as ensured by the PT-symmetry property. The system $H_3$ is non-PT-symmetric and both its eigenfunction and eigenvalue are complex. However, if we treat $x_1$, $x_2$, and $x_3$ as complex variables, the above distinctions between $H_1$, $H_2$, and $H_3$ no longer exist, because now their eigenfunctions are all complex and can be made identical by complex coordinate translation as indicated in Eq. (4.4), and meanwhile their eigenvalues can be made coincident by constant shift of the complex energy as shown in Eq. (4.6) and Eq. (4.7).

It can be further shown that the eigen-trajectories for $H_1$, $H_2$, and $H_3$ are also shift-invariant. Like Eq. (3.6), the eigen- trajectories for $H_2$, and $H_3$ are governed by the following equations of motion:

$$\frac{dx_2}{dt} = -\mathrm{i}\frac{d\ln\psi_{2,n}(x_2)}{dx_2},$$
$$\frac{dx_3}{dt} = -\mathrm{i}\frac{d\ln\psi_{3,n}(x_3)}{dx_3}, \quad (4.8)$$

where $\psi_{2,n}(x_2)$ and $\psi_{3,n}(x_3)$ are given by Eqs.(4.3). Especially, for $n=0$ we have,

$$\frac{dx_2}{dt} = \mathrm{i}\left(x_2 + \frac{\mathrm{i}}{2}\right), \quad \frac{dx_3}{dt} = \mathrm{i}\left(x_3 + \frac{\mathrm{i}}{2} - \frac{1}{2}\right). \quad (4.9)$$

In comparison with Eqs. (3.7) and (4.4), it can be seen that the eigen-trajectories $x_1(t)$, $x_2(t)$, and $x_3(t)$ obey a shift-invariant relation

$$x_1(t) = x_2(t) + \mathrm{i}/2, \quad (4.10\mathrm{a})$$

$$x_1(t) = x_3(t) + \mathrm{i}/2 - 1/2. \quad (4.10\mathrm{b})$$

An illustration of this shift-invariant property for trajectories is shown in Fig.3, where we can see that the eigen-trajectory $x_2(t)$ can be coincided with $x_1(t)$ by a parallel translation $-\mathrm{i}/2$ according to Eq. (4.10a). Similarly, Fig.4 shows a parallel translation $-\mathrm{i}/2+1/2$ between $x_1(t)$ and $x_3(t)$.

Apparently, the shift-invariant property for complex trajectory reveals the connection between Hermitian, PT-symmetric and non-PT- symmetric systems. The boundary between those Hamiltonians vanishes if we view the particle motion from complex space.

## 5. Linear-Mapping Invariants of Eigen- Structures

In the preceding section, we have seen that the quantum trajectories of the three Hamiltonians $H_1$, $H_2$ and $H_3$ corresponding to the three variations of harmonic oscillator have the same shape and can be made identical by linear translation as indicated by Eq. (4.10). This observation inspires our interest in searching for the class of quantum systems that may share the same eigenfunctions and quantum trajectories under linear mapping. Consider the following 1D classical Hamiltonian with an arbitrary potential function $V(x_1)$:

$$H_1^{(c)} = \frac{1}{2m} p_1^2 + V(x_1), \quad x_1, \ p_1 \in \mathbb{C}. \tag{5.1}$$

The eigenvalue problem $H_1 \psi_1 = E_1 \psi_1$ reads

$$\left[-\frac{\hbar^2}{2m} \frac{d^2}{dx_1^2} + V(x_1)\right] \psi_1(x_1) = E_1 \psi_1(x_1). \tag{5.2}$$

It is emphasized that the above Schrödinger equation is defined in a complex domain, having complex eigenfunction $\psi_1(x_1)$ and complex eigenvalue $E_1$. The quantum Hamiltonian associated with an eigenstate $\psi_1(x_1)$ satisfying Eq. (5.2) is given by

$$H_1 = \frac{1}{2m} p_1^2 + V(x_1) - \frac{\hbar^2}{2m} \frac{d^2 \ln \psi_1(x_1)}{dx_1^2}, \tag{5.3}$$

and the related quantum trajectories $x_1(t)$ are governed by the differential equation

$$\frac{dx}{dt} = \frac{p_1}{m} = \frac{\hbar}{im} \frac{d \ln \psi_1(x_1)}{dx_1}. \tag{5.4}$$

Our aim is to find out what type of variation made to $H_1^{(c)}$ will not change the shape of the quantum trajectories described by Eq. (5.4). A class of Hamiltonians meeting this requirement assumes the following form

$$H^{(c)} = \frac{1}{2ma^2} p^2 + V(ax + b) + c, \tag{5.5}$$

where $a \neq 0$, $b$ and $c$ are arbitrary complex constants. The Schrödinger equation accompanying the Hamiltonian in Eq. (5.5) reads

$$\left[-\frac{\hbar^2}{2ma^2} \frac{d^2}{dx^2} + V(ax + b) + c\right] \psi = E\psi. \tag{5.6}$$

Through the linear mapping

$$x_1 = ax + b, \tag{5.7}$$

Eq. (5.6) can be transformed to

$$\left[-\frac{\hbar^2}{2m}\frac{d^2}{dx_1^2} + V(x_1)\right]\psi(x_1) = (E - c)\psi(x_1). \tag{5.8}$$

Comparing Eq.(5.8) with Eq.(5.2), we obtain the eigenfunction $\psi(x)$ and the eigenvalue $E$ for the Schrödinger equation (5.8) as

$$\psi(x) = \psi_1(ax + b), \quad E = E_1 + c, \tag{5.9}$$

which implies that the class of the Hamiltonian operators $H$ parameterized by the three free parameters $a$, $b$ and $c$ in Eq. (5.5) shares the same eigenfunctions with $H_1$ under the linear mapping (5.7). This property of wavefunction invariance under linear conformal mapping is valid for any potential $V(x)$. The preceding section has considered several examples for a quadratic potential $V(x) = x^2/2$.

Besides the wavefunctions, all the quantum trajectories in the class of the quantum systems described by Eq. (5.5) are also linear-mapping invariant. The quantum Hamiltonian associated with $H^{(c)}$ is given by

$$H = \frac{p^2}{2ma^2} + V(ax + b) + c - \frac{\hbar^2}{2m}\frac{d^2 \ln \psi(x)}{dx^2} \tag{5.10}$$

and the quantum trajectory in a state $\psi(x)$ satisfying Eq. (5.6) is governed by the differential equation:

$$\frac{dx}{dt} = \frac{\partial H}{\partial p} = \frac{p}{ma^2} = \frac{1}{ma^2}\frac{\hbar}{i}\frac{d \ln \psi(x)}{dx}. \tag{5.11}$$

Substituting the relation (5.9), we have

$$a\frac{dx}{dt} = \frac{1}{m}\frac{\hbar}{ia}\frac{d \ln \psi_1(ax + b)}{dx}, \tag{5.12}$$

which, via the linear mapping $x_1 = ax + b$, becomes

$$\frac{dx_1}{dt} = \frac{\hbar}{\mathrm{i}m} \frac{d \ln \psi_1(x_1)}{dx_1}. \tag{5.13}$$

This is just the equation of motion (5.4) for the base quantum system $H_1$. Therefore, all the quantum trajectories $x(t)$ in the class of the quantum systems described by Eq. (5.5) can be produced by the base trajectory $x_1(t)$ via the inverse linear mapping

$$x(t) = \frac{1}{a} x_1(t) - b , \tag{5.14}$$

where the factor $b$ represents a linear shift of the trajectory $x_1(t)$ in the complex plane and the factor $1/a$ represents a combined effect of rotation and magnification. We thus have constructed a class of quantum systems that are invariant under linear mapping. Symbolically, this class of linear-mapping invariant quantum systems can be parameterized by

$$\left\{ H \middle| H = \frac{-1}{2a^2} \frac{d^2}{dx^2} + V(ax+b) + c, \ \forall a,\ b,\ c \in \mathbb{C} \right\} \tag{5.15}$$

Different combinations of $a$, $b$ and $c$ will lead to different quantum systems; some have Hermitian symmetry (both eigenvalues and eigenfunctions are real), some have PT symmetry but not Hermitian symmetry (Eigenvalues are real but eigenfunctions are complex), and some have broken PT symmetry (both eigenvalues and eigenfunctions are complex). The results obtained in the preceding section are only some special cases of shift invariance applied to the quadratic potential $V(x) = x^2/2$. Let us examine how the results of the preceding section can be fitted to the present framework. The base Hamiltonian to be considered is

$$H_1^{(c)} = (p_1^2 + x_1^2)/2. \tag{5.16}$$

The eigenfunctions $\psi_1(x_1)$ and the eigenvalues $E_1$ of $H_1$ are given by Eq. (3.3) and Eq. (3.4). In the following, we consider several varieties of $H_1$ and study their invariants with respect to $H_1$.

**(A) Shift-invariant quantum systems**

Two quantum systems are shift invariant, if their quantum trajectories can be connected by the linear mapping (5.7) with $a = 1$. The first variation of $H_1^{(c)}$ considered in the preceding section is $H_2^{(c)}$:

$$H_2^{(c)} = (p_2^2 + x_2^2 + \mathrm{i} x_2)/2 = p^2/2 + (x_2 + \mathrm{i}/2)^2/2 + 1/8$$
$$= p_2^2/2 + V(x_2 + \mathrm{i}/2) + 1/8, \quad (5.17)$$

where we have $a = 1$, $b = \mathrm{i}/2$, and $c = 1/8$ by comparing to the standard form (5.5). Hence, via the linear mapping $x_1 = x_2 + \mathrm{i}/2$, the eigenfunctions for the Hamiltonian $H_2$ is obtained as

$$\psi_2(x_2) = \psi_1(x_1) = \psi_1(x_2 + \mathrm{i}/2) = C_n H_n(x_2 + \mathrm{i}/2) e^{-(x_2 + \mathrm{i}/2)^2/2}. \quad (5.18)$$

It can be seen that $\psi_2(x_2)$ is a complex function of $x_2$, while its eigenvalues are still real

$$E_{2,n} = E_{1,n} + c = n + 5/8. \quad (5.19)$$

The corresponding eigen-trajectory $x_2(t)$ is given by Eq. (5.14) as

$$x_2(t) = (x_1(t) - b)/a = x_1(t) - \mathrm{i}/2. \quad (5.20)$$

Since the base eigen-trajectories $x_1(t)$ at $n = 0$ are concentric circles with center at the origin, $x_2(t)$ represents down-shift concentric circles with center at $-\mathrm{i}/2$, as shown in Fig. 3.

The second variation of $H_1^{(c)}$ considered in the preceding section is $H_3^{(c)}$:

$$H_3^{(c)} = (p_3^2 + x_3^2 - x_3 + \mathrm{i} x_3)/2 = \frac{1}{2} p_3^2 + \frac{1}{2}(x_3 + \mathrm{i}/2 - 1/2)^2 + \frac{\mathrm{i}}{4}, \quad (5.21)$$

where the three constants are identified as

$$a = 1,\ b = \mathrm{i}/2 - 1/2,\ c = \mathrm{i}/4. \quad (5.22)$$

Employing the linear mapping $x_1 = ax_3 + b = x_3 + i/2 - 1/2$ and repeating the above evaluation process, we obtain the eigenfunction, eigenvalue and eigen-trajectory for $H_3$, respectively, as

$$\psi_3(x_3) = \psi_1(x_1) = \psi_1(x_3 + i/2 - 1/2) = C_n H_n(x_3 + i/2 - 1/2)e^{-(x_3+i/2-1/2)^2/2}$$

$$E_3 = E_1 + c = n + 1/2 + i/4,$$

$$x_3(t) = (x_1(t) - b)/a = x_1(t) - i/2 + 1/2.$$

Even though both the eigenvalues and eigenfunctions of $H_3$ are complex, its eigen-trajectories can be made identical, by linear shift, to those of $H_1$ whose eigenvalues and eigenfunctions are real. We say that $H_2$ and $H_3$ are shift invariants of $H_1$, because their quantum trajectories can be made coincident with those of $H_1$ by linear shift.

**(B) Rotation-invariant quantum systems**

Two quantum systems are said to have rotation invariant, if their quantum trajectories can be made identical by relative rotation; in other words, they are connected by the mapping (5.7) with $a = e^{i\theta}$, $\theta \neq 0$. A typical rotation-invariant Hamiltonian with respect to $H_1^{(c)}$ is given by

$$H_4^{(c)} = -\frac{1}{2}p_4^2 - \frac{1}{2}x_4^2 = \frac{1}{2i^2}p_4^2 + \frac{1}{2}(ix_4)^2 \quad (5.23)$$

which gives the values of the three parameters as

$$a = i = e^{\pi i/2}, \ b = c = 0. \quad (5.24)$$

With the mapping $x_1 = ax_4 + b = e^{\pi i/2}x_4$, the eigenfunctions and eigenvalues of $H_4$ can be found as

$$\psi_4(x_4) = \psi_1(ix_4) = C_n H_n(ix_4)e^{-(ix_4)^2/2} \quad (5.25a)$$

$$E_4 = E_1 + c = E_1 = n + 1/2. \tag{5.25b}$$

Comparing Eq. (5.23) with Eq. (5.16), the Hamiltonian $H_4$ can be conceived of as describing the motion of a unit negative mass under the action of a negative quadratic potential $V(x) = -x^2/2$. Equation (5.25b) implies that the total energy $E_4$ of this negative-mass system is positive real and is equal to the total energy $E_1$ of a unit positive mass subjected to a positive quadratic potential $V(x) = x^2/2$. The eigenfunctions of this negative-mass system are unbound, as shown in Eq.(5.25a). However, unbound eigenfunctions do not imply unbound quantum trajectory, because the quantum trajectories of $H_4$ are simply obtained from the bound trajectories $x_1(t)$ via a 90-degree clockwise rotation:

$$x_4(t) = e^{-\pi i/2} x_1(t). \tag{5.26}$$

Except for this $90°$ phase shift, the quantum trajectory of a negative-mass particle subjected to a negative potential $-V$ is identical to the quantum trajectory of a positive-mass particle subjected to a positive potential $V$. This point can be confirmed by the equation of motion (5.11) with $a = i$ and $\hbar = m = 1$:

$$\frac{dx_4}{dt} = i \frac{d}{dx_4} \ln H_n(ix_4) e^{x_4^2/2}$$

In the ground state $H_0(ix_4) = 1$, the above equation reduces to

$$\dot{x}_4 = ix_4, \tag{5.27}$$

which is identical to the equation of motion (3.7) for the Hamiltonian $H_1$ with positive mass.

**(C) Rotation-Shift invariant systems**

This class of quantum systems, which are described by Eq. (5.5) with $a = e^{i\theta}$, $\theta \neq 0$ and $b \neq 0$, can be made identical to $H_1$ by a combined rotation and shift operation. One of the systems in this class takes on the following form:

$$H_5^{(c)} = -\frac{1}{2}p_5^2 - \frac{1}{2}x_4^2 + \frac{i}{2}x_4 + \frac{3}{8} = \frac{1}{2i^2}p_5^2 + \frac{1}{2}(ix_5 + i/2)^2 + \frac{1}{2}, \quad (5.28)$$

corresponding to the parameter setting

$$a = i, \ b = i/2, \ c = 1/2. \quad (5.29)$$

The quantum trajectory $x_5(t)$ is a rotation-shift invariant of the base trajectory $x_1(t)$; they are related by the linear mapping (5.14): $x_5(t) = (x_1(t) - b)/a = e^{-\pi i/2}x_1(t) - 1/2$. It means that the trajectory $x_5(t)$ is obtained from $x_1(t)$ by a $90°$ clockwise rotation followed by a $1/2$ shift to the left, as shown in Fig.5 for the first excited state $n = 1$. The equation of motion for $x_5(t)$ is given by

$$\frac{dx_5}{dt} = -i\frac{(4x_5^2 + 4x_5 + 5)}{2(2x_5 + 1)}, \quad (5.30)$$

whose equilibrium points are at $-1/2 \pm i$ that are the transformation of the equilibrium points $\pm 1$ of $x_1(t)$ via the aforementioned rotation-shift operations. Although quantum system $H_5$ possesses bound eigen-trajectory $x_5(t)$ and real positive eigenvalues $E_5 = E_1 + 1/2 = n + 1$, its eigenfunctions are unbound and complex:

$$\psi_5(x_5) = C_n H_n(ix_5 + i/2)e^{(x_5 + 1/2)^2/2}. \quad (5.31)$$

**(D) Stretch invariant quantum systems**

The stretch effect is controlled by the parameter $a$ in Eq. (5.5); a quantum trajectory $x(t)$ is expanded from the base trajectory $x_1(t)$ if $|a| < 1$, and is contracted if $|a| > 1$. A representative example is given by

$$H_6^{(c)} = \frac{1}{8}p_6^2 + 2x_6^2 + 4x_6 + \frac{5}{2} = \frac{1}{2}\frac{p_6^2}{2^2} + \frac{1}{2}(2x_6 + 2)^2 + \frac{1}{2}, \quad (5.32)$$

which is characterized by the three parameters

$$a = 2,\ b = 2,\ c = 1/2.  \tag{5.33}$$

The quantum trajectory $x_6(t)$ of $H_6$ is related to the base trajectory $x_1(t)$ via the linear mapping

$$x_6(t) = x_1(t) - b\ /a = x_1(t)/2 - 1,  \tag{5.34}$$

which amounts to a $1/2$ contraction and a 1-unit shift to the left, as shown in Fig.6.

A real $a$ has a pure magnification effect, while a complex $a$ owns a combined magnification and rotation effect. The role of the parameter $a$ as a magnification factor of quantum trajectory is attributed to its influence on the particle's mass. From Eq. (5.5) we can see that the system described by Eq. (5.5) has an equivalent mass as

$$m_{eq} = ma^2  \tag{5.35}$$

Hence, if $|a| < 1$, we have a reduced mass and an enlarged trajectory; conversely, if $|a| > 1$, we have an enlarged mass and a reduced trajectory. Moreover, when $a$ is a pure imaginary number, the equivalent mass becomes negative. In the quantum system $H_4$ described by Eq. (5.23), we have encountered a quantum system with negative mass and negative potential, which behaves like a positive mass under the action of a positive potential. In the most general case, we may have a complex equivalent mass $m_{eq} \in \mathbb{C}$ if $a$ is a complex number, which corresponds to a linear mapping with a combined effect of magnification and rotation. In short, the class of quantum systems characterized by Eq. (5.15) collect all the systems having positive, negative and complex masses whose eigenfunctions and quantum trajectories can be made identical by linear mapping, i.e., via the operations of translation, rotation and magnification.

## 6. Complex Energy

It might be unusual to see a complex potential in the conventional quantum mechanics. However, complex potentials have been wildly used in quantum field theory and quantum statistics. Especially, the related study to non-Hermitian Hamiltonian in those theories can be found in a local self-interacting scalar quantum field study [34], and in discussing statistical properties of eigenvectors of non-Hermitian random matrices [35]. On the other hand, there are some physical realizations of complex external potentials: delocalization issues in superconductors and population biology [36,37]; degenerate piezoelectric properties for some special materials [38]; dissipative process description [39] and quantum transport [40], and so on.

In the previous discussions, we pointed out the relationship between a complex Hamiltonian and its accompanied eigenvalues. The results indicate that a quantum system with complex Hamiltonian may possess complex eigenvalues. It is natural for a particle to have complex energy spectrum when it suffers a complex-valued potential. For example, the complex Hamiltonian in Eq. (4.2) gives us a complex-valued energy spectrum indicated by Eq.(4.7b).

There are at least three sources of complex energy. The first source comes from the quantum systems with complex eigenvalues, as discussed in the previous sections. The second possible source of the complex energy originates from the entanglement of eigenstates with real eigenvalues [26]. In a state transition process, both the initial and final energy levels are real but the intermediate energy level may be complex. Under the framework of quantum Hamilton mechanics, a continuous transition trajectory connecting two real eigenstates can be traced in the complex plane and a continuous complex-energy transfer can be indicated. For example, consider a state transition between the $n=0$ state and the $n=1$ state of a harmonic oscillator, for which the wavefunction is described as:

$$\Psi_E = \begin{cases} \psi_0 = e^{-x^2/2} e^{-iE_0 t} & t \leq 0 \\ (1-t)e^{-x^2/2} + 2txe^{-x^2/2}, & 0 < t < 1 \\ \psi_1 = 2xe^{-x^2/2} e^{-iE_1 t} & t \geq 1 \end{cases} \qquad (6.1)$$

where $E_0 = 1/2$ and $E_1 = 3/2$ denote the energy level for the initial state and final state. The governing equations for this state transition process can be derived from Eq. (2.9) as

$$p = \frac{dx}{dt} = \begin{cases} \mathrm{i}x & t \leq 0 \\ \mathrm{i}x - \mathrm{i}\dfrac{2t}{2tx - t + 1}, & 0 < t < 1 \\ \mathrm{i}(x^2 - 1)/x & t \geq 1 \end{cases} \qquad (6.2)$$

Complex trajectories in the time ranges $t \leq 0$ and $t \geq 1$ are just eigen-trajectories for the $n=0$ state and the $n=1$ state, as shown in Fig.1 and Fig.2. The governing equation in the time interval $0 < \tau < 1$ is a non-autonomous nonlinear differential equation whose solution can only be found numerically. Connecting the complex trajectories $x(t) = x_R(t) + \mathrm{i} x_I(t)$ for the above three time ranges offers a continuous manifestation of the entire state transition process from the $n=0$ state to the $n=1$ state. The total energy $E_{\text{Total}}$ is observed to evolve continuously from $E_{\text{Total}} = E_0 = 1/2$ to $E_{\text{Total}} = E_1 = 3/2$ via the complex trajectory. The overall time history of $E_{\text{Total}}(t)$ can be described by:

$$E_{\text{Total}} = p^2(t)/2 + x^2(t)/2 + Q(x,t)$$
$$= \begin{cases} 1/2 & t \leq 0 \\ \dfrac{1}{2}\dfrac{6tx(t) - t + 1}{2tx(t) - t + 1}, & 0 < t < 1 \\ 3/2 & t \geq 1 \end{cases} \quad (6.3)$$

where the momentum $p$ and the quantum potential $Q$ are given, respectively, by Eq. (6.2) and Eq. (2.5). It can be seen that the total energy $E_{\text{Total}}(t)$ is complex-valued due to the complex feature of $x(t)$. The evolution of $E_{\text{Total}}(t)$ on the complex plane is demonstrated in Fig.7, showing that except at $t=0$ and $t=1$, the particle during the transition process does have complex total energy.

The third source of complex energy stems from a general quantum state that is neither a complex eigenstate nor an entanglement of real eigenstates. In fact, the condition of having a particle in an eigenstate is very stringent. It has to satisfy two requirements: (1) the particle's total energy $E_{\text{Total}}$ has to be one of the eigen-energies of the system, and (2) the particle's initial velocity $\dot{x}_0$ should be exactly equal to:

$$\dot{x}_0 = \frac{1}{\mathrm{i}} \frac{d \ln \psi_n(x)}{dx}\bigg|_{x=x_0}, \quad (6.4)$$

where $\psi_n(x)$ is one of the eigenfunctions of the system. It is apparent that in a scattering problem the initial velocity of an incident particle may not happen to be the one given by Eq. (6.4), nor may its total energy be one of the eigen-energy. In other words, scattering trajectories in general do not coincide with eigen-trajectories.

Furthermore, the quantum function describing a scattering particle is often unbound to allow the particle to approach to infinity.

The total energy accompanied with a general unbound state $\psi(x)$ is given by Eq. (2.12) as

$$E_{\text{Total}} = \frac{m}{2}\dot{x}^2 + V(x) - \frac{\hbar^2}{2m}\frac{d^2\psi(x)}{dx^2}. \tag{6.5}$$

Noting that the value of $E_{\text{Total}}$ is determined contingently by incident condition $\dot{x}_0 \in \mathbb{C}$ and $x_0 \in \mathbb{C}$, the total energy $E_{\text{Total}}$ for a scattering particle is in general a complex number [41].

In literature, the complex potential plays a central role in scattering issues, such as high-energy proton-proton scattering [42], $^2\Sigma_u^+$ shape resonance state of $H_2^-$ at the self-consistent-field level [43], and arrival times discussions [44,45], and so on. In addition, complex absorbing potentials and complex scaling could provide important tools in stationary or time dependent scattering calculations [46], in scattering interferences [47], and in determining the resonance energy and width [48,49], and so on. In a recent research, quantum scattering trajectories on a complex plane were discussed on the basis of quantum Hamilton-Jacobi formalism [50], where compatible results with our complex energy discussions were proposed from the scattering standpoint.

## 7. Conclusions

The existence of complex Hamiltonian and complex energy is not an assumption but is derivable from the Schrödinger equation. By writing the Schrödinger equation in the form of the Hamilton-Jacobi equation, we prove that an intrinsic complex Hamiltonian and its related complex Hamilton equations of motion arise naturally within every quantum system. Solving complex Hamilton equations of motion then leads to complex-valued position and momentum. The conventional real-valued quantum mechanics is found to be a special case of the complex-valued quantum mechanics, where quantum motion is evaluated along the real axis.

When viewed from the complex domain, the boundary between Hermitian, PT-symmetric, and non-PT- symmetric systems disappears; they can be made identical by linear complex mapping. It is shown that the eigen-structures of a complex-valued quantum system, such as its eigenfunctions, eigenvalues and eigen- trajectories, are invariant under linear complex transformation. A class of quantum systems that share the same eigen-structures has been characterized in this paper.

Having complex energy is a common property of complex-valued quantum systems. A complex energy may originate from an eigenstate with complex energy spectrum, from an entanglement of real eigenstates, or from a scattering state with contingently assigned initial position and momentum. Discussions and analyses of complex spectrums are essential for the applications in the field of superconductor, semiconductor, and biology, and other scientific usages.

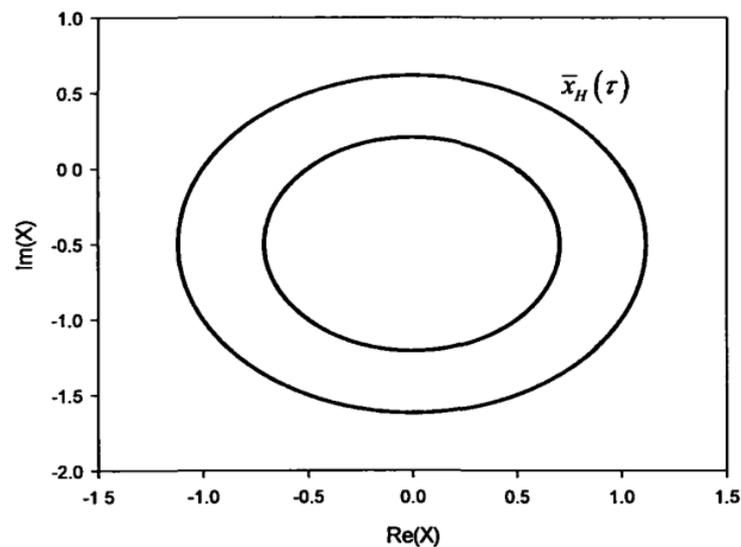

**Fig. 1:** Complex trajectories in the ground state of a harmonic oscillator

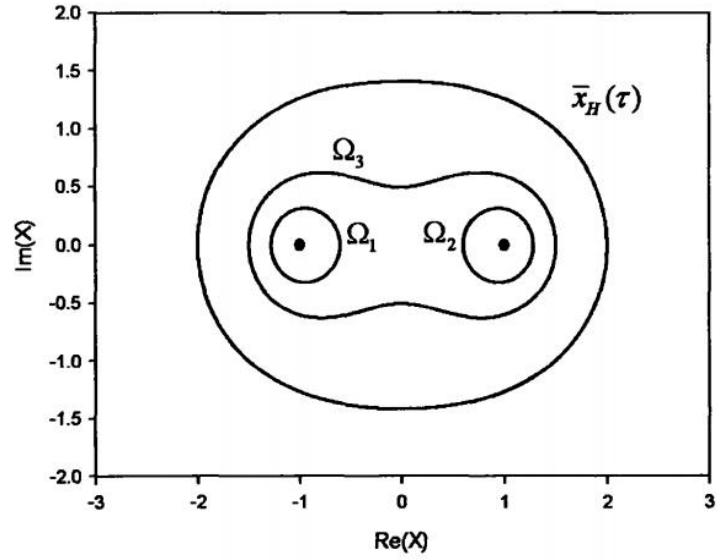

**Fig.2:** Complex trajectories in the first excited state of a harmonic oscillator.

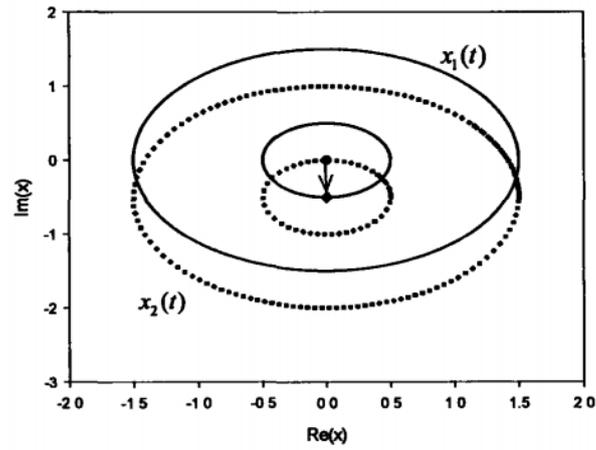

**Fig.3:** The parallel translation from $x_1(t)$ to $x_2(t)$ by the relation (4.10a)

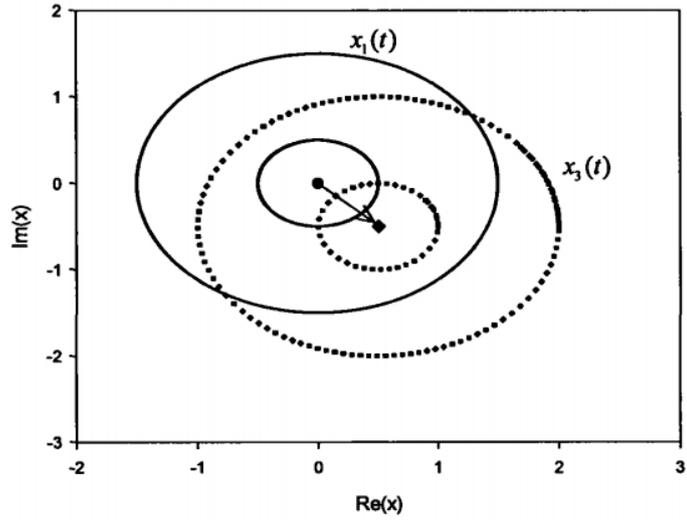

**Fig.4:** The parallel translation from $x_1(t)$ to $x_3(t)$ by relation (4.10b)

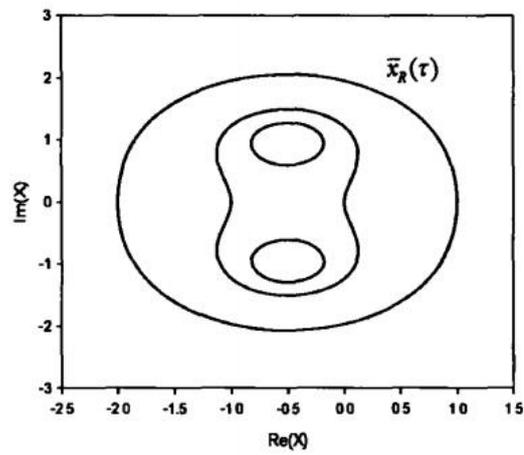

**Fig.5:** The eigen-trajectories $x_5(t)$ have $90^0$ clockwise rotation and $-1/2$ shift along the $x$ axis with respect to the base eigen-trajectories $x_1(t)$ indicated in Fig.2.

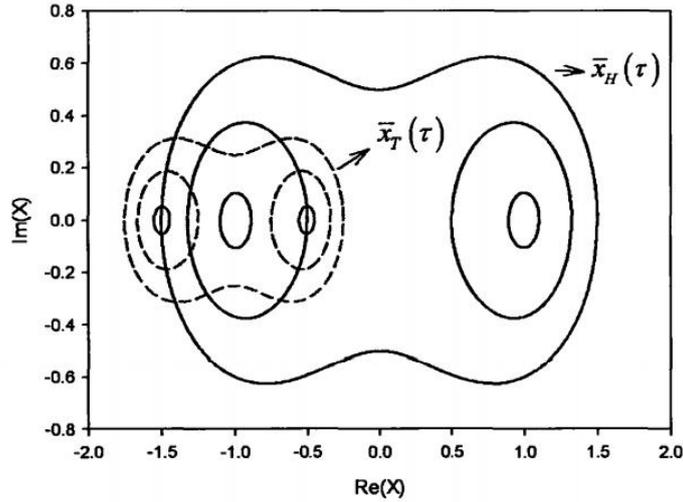

**Fig.6:** The eigen-trajectories $x_6(t)$ are obtained from the base eigen-trajectories $x_1(t)$ by a 1/2 contraction and a 1-unit shift to the left.

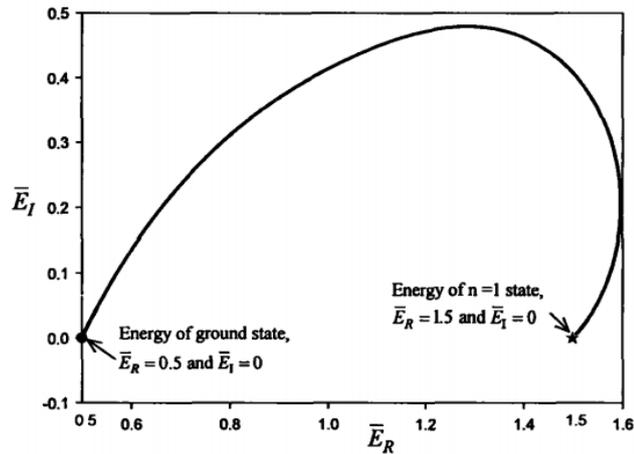

**Fig.7:** The complex-energy transition history for a state-transition process starting from the initial state $n=0$ to the terminal state $n=1$.